\documentclass[11pt,twoside]{article}

\usepackage{asp2006}
\usepackage{epsf}
\usepackage{psfig}
\usepackage{lscape}

\begin{document}

\title{The formation of very wide binaries}   

\author{M.B.N.~Kouwenhoven\altaffilmark{1,2},
S.P. Goodwin\altaffilmark{2}, Richard J.~Parker\altaffilmark{2},
M.B.~Davies\altaffilmark{3}, D.~Malmberg\altaffilmark{3} and
P.~Kroupa\altaffilmark{4}} 

\altaffiltext{1}{Gruber Foundation Fellow at the Kavli Institute for
Astronomy and Astrophysics, Peking University, Yi He Yuan Lu 5, Hai
Dian Qu Beijing 100871, P.\,R. China}
\altaffiltext{2}{University of Sheffield, Hicks Building, Hounsfield
Road, Sheffield S3\,7RH, United Kingdom}
\altaffiltext{3}{Lund Observatory, Box\,43, SE-221\,00, Lund, Sweden}
\altaffiltext{4}{Argelander Institute for Astronomy, University of
Bonn, Auf dem H\"{u}gel 71, 53121, Bonn, Germany}


\begin{abstract} 
Over the last decades, numerous wide ($>1000$ AU) binaries have been
discovered in the Galactic field and halo. The origin of these wide
binaries cannot be explained by star formation or by dynamical
interactions in the Galactic field. We explain their existence by wide
binary formation during the dissolution phase of young star
clusters. In this scenario, two single stars that leave the dissolving
cluster at the same time, in the same direction, and with similar
velocities, form a new, very wide binary. Using $N$-body simulations
we study how frequently this occurs, and how the orbital parameters of
such binaries depend on the properties of the cluster from which they
originate. The resulting wide binary fraction for individual star
clusters is $1-30$\%, depending on the initial conditions. As most
stars form as part of a binary or multiple system, we predict that a
large fraction of these wide ``binaries'' are in fact wide triple and
quadruple systems.
\end{abstract}



\section{Observations and origin of wide binaries}

Observations have indicated that the majority of the stars form in
star clusters \citep[e.g.,][]{tk_ladalada} and as part of a binary
system \citep[e.g.,][]{tk_duquennoy1991}. Surveys for binarity have
indicated that a significant number of binaries have orbital
separations that are comparable with the typical size of star clusters
in which they are thought to have formed.  These wide binaries are
usually identified as common proper motion pairs \citep[e.g.,][and
numerous others]{tk_wasserman, tk_chaname, tk_lepine, tk_makarov,
tk_quinn} or by employing statistical methods \citep[e.g.,][Longhitano
\& Binggeli 2009, submitted]{tk_bahcall, tk_garnavich, tk_gould}. The
wide binary fraction in the separation range $10^3~\mbox{AU} < a <
0.1$~pc is of order 15\% \citep{tk_duquennoy1991,tk_poveda2007}, with
a drop-off at around $0.1-0.2$~pc, which corresponds to the stability
limit for wide binaries in the Galactic field. In the Galactic halo,
where the stellar density is smaller, wide binaries with separations
of up to several parsecs are known \citep{tk_chaname}. The origin of
these wide binaries in the field and halo has long been a
mystery. Below, we discuss four potential formation theories of wide
binaries:
\begin{itemize}
\item {\bf Clustered star formation}. This mode of wide binary
formation is excluded, given the compact nature of the embedded
clusters ($< 1$~pc). Even if it were possible to form such a wide
binary via the star formation process, it would immediately be
destroyed by dynamical interactions with other cluster stars. Hence,
clustered star formation is not a viable option.

\item {\bf Diffuse star formation}. Although this may be a possible
formation mechanism for some wide binaries, diffuse (or isolated) star
formation is rare, and it is therefore difficult to account for a the
large number of wide binary systems in the Galactic field.

\item {\bf Dynamical capture}. Binary formation by dynamical capture
is theoretically possible. This mechanism requires a third star to
remove kinetic energy from the system, such that the binding energy of
the two remaining stars becomes negative. It requires the three stars
to be at the same place at the same time, with fine-tuned velocities
and impact angles. In the Galactic field, this combination of
parameters is extremely unlikely, given the stellar density and
velocity dispersion in the field \citep{tk_goodman1993}. Therefore,
this possibility of forming wide binaries is ruled out. Note that
binary formation by dynamical capture is possible in star clusters, as
the density is higher and the velocity dispersion lower. However, this
does not result in {\em wide binaries}, as these are immediately
destroyed after their formation in young, dense star clusters.

\item {\bf Formation during star cluster dissolution}. The vast
majority of star clusters are short-lived, and dissolve into the field
star population quickly after their formation. Two unbound stars that
are close in phase-space may form a wide binary system when a young
star cluster dissolves.  We have tested this possibility using
$N$-body simulations, and find that this process results in a wide
binary fraction of 1--30\%, depending on the initial conditions of the
star cluster (Kouwenhoven et al., 2009, in prep).  Wide binary
formation during the star cluster dissolution phase is therefore a
viable mechanism for the formation of the observed wide binary
population.

\end{itemize}

\section{$N$-body simulations and results}


Our hypothesis that a significant number of wide binaries form during
the star cluster dissolution process, can easily be tested using
$N$-body simulations.  We therefore carry out star cluster simulations
using the STARLAB package \citep{tk_starlab}, and study how the
resulting wide binary population depends on the initial conditions of
a star cluster.  We draw $N$ single stars from the \cite{tk_kroupa}
mass distribution in the mass range $0.1M_\odot < M < 50M_\odot$. We
perform simulations of clusters starting out from a \cite{tk_plummer}
morphology, and substructured clusters with a fractal dimension of
1.5, following the prescriptions of \cite{tk_goodwin2004}. By varying
the initial cluster mass, size, and morphology, we study how the
properties of the wide binary population depend on these, and obtain
the following results:
\begin{enumerate}
\item The binary fraction among the dissolved stellar population ranges
between $F=1\%$ and 30\%, depending on the cluster properties. The binary
fraction, in this case, is measured as $F =
(B+T+\dots)/(S+B+T+\dots)$, where $S$, $B$, and $T$ denote the number
of single stars, binary systems, and triple systems in the cluster.

\item More massive star clusters result in smaller binary fractions
than low-mass clusters. Clusters with a spherical, smooth stellar
density distribution form fewer wide binaries than substructured
clusters of the same size and mass.

\item The typical semi-major axis $a$ of a newly formed binary is
comparable to the initial size $R$ of the star cluster in which it was
born (the reason for this being that $R$ is the only size scaling that
is present in the initial cluster). The resulting semi-major axis
distribution is generally bimodal, consisting of a {\em dynamical
peak} with binaries formed by dynamical interactions in the cluster
centre, and a {\em dissolution peak} with binaries formed during the
cluster dissolution phase.

\item The formation of wide binaries during the star cluster
dissolution phase is a random process, resulting in a thermal
eccentricity distribution, a (gravitationally-focused) randomly-paired
mass ratio distribution, and random alignments between the orbital
spin vector and the stellar rotation axes.

\item Star clusters with a non-zero primordial binary fraction form
wide triple and quadruple systems, i.e., the two components that make
up the newly formed wide ``binary'' are in fact primordial binaries,
rather than single stars. The ratio of triple to quadruple systems
among very wide orbits is therefore indicative of the primordial
binary fraction. Given that the primordial binary fraction is large,
we predict a high frequency of triple and quadruple systems among the
known wide ``binary'' systems.
\end{enumerate}

\section{Summary}

Approximately 15\% of the binary systems in the Galactic field have
separations in the range $10^3~\mbox{AU} < a < 0.1$~pc. Their origin
cannot be explained by star formation or by dynamical capture in the
Galactic field.  We have carried out $N$-body simulations to test the
hypothesis that these wide binaries are formed during the dissolution
process of young star clusters, where two stars happen to fly off in
the same direction and form a new, wide binary (Kouwenhoven et al., 2009, in prep).

Our simulations indicate a binary fraction of $1-30\%$ in the
separation range $10^3~\mbox{AU} < a < 0.1$~pc, depending on the
initial conditions of the star clusters. Clusters with more
substructure and with a smaller cluster mass result in a higher wide
binary fraction. The resulting separation distribution is generally
bimodal, and has a {\em dynamical peak} of close binaries formed via
dynamical interactions in the cluster, and a {\em dissolution peak}
consisting of wide binaries formed during the dissolution process of
the star cluster.

For star clusters with a high primordial binary fraction, these
primordial binaries pair up into wide systems. Given that most stars
form as part of a binary system, we predict that most observed wide
``binaries'' are in fact wide triple and quadruple systems. The ratios
between wide binary, triple, and quadruple systems are therefore
indicative of the primordial binary fraction.

\acknowledgements MBNK was supported by the Peking University Hundred
Talent Fund (985), the Peter and Patricia Gruber Foundation through
the PPGF fellowship, and by PPARC/STFC under grant number
PP/D002036/1. RJP acknowledges support from STFC.



\begin{thebibliography}{}

\bibitem[Bahcall \& Soneira(1981)]{tk_bahcall} Bahcall, J.~N., \&
Soneira, R.~M.\ 1981, \apj, 246, 122

\bibitem[Chanam{\'e} \& Gould(2004)]{tk_chaname} Chanam{\'e}, J., \&
Gould, A.\ 2004, \apj, 601, 289

\bibitem[Duquennoy \& Mayor(1991)]{tk_duquennoy1991} Duquennoy, A., \&
Mayor, M.\ 1991, \aap, 248, 485

\bibitem[Garnavich(1988)]{tk_garnavich} Garnavich, P.~M.\ 1988, \apjl,
335, L47

\bibitem[Goodman \& Hut(1993)]{tk_goodman1993} Goodman, J., \& Hut, P.\
1993, \apj, 403, 271

\bibitem[Goodwin \& Whitworth(2004)]{tk_goodwin2004} Goodwin, S.~P.,
\& Whitworth, A.~P.\ 2004, \aap, 413, 929

\bibitem[Gould et al.(1995)]{tk_gould} Gould, A., Bahcall, J.~N.,
Maoz, D., \& Yanny, B.\ 1995, \apj, 441, 200

\bibitem[Kouwenhoven et al.(2005)]{tk_2005} Kouwenhoven, M.~B.~N.,
Brown, A.~G.~A., Zinnecker, H., Kaper, L., \& Portegies Zwart, S.~F.\
2005, \aap, 430, 137

\bibitem[Kouwenhoven et al.(2007)]{tk_2007} Kouwenhoven, M.~B.~N.,
Brown, A.~G.~A., Portegies Zwart, S.~F., \& Kaper, L.\ 2007, \aap,
474, 77

\bibitem[Kroupa(2001)]{tk_kroupa} Kroupa, P.\ 2001, \mnras, 322, 231

\bibitem[Lada \& Lada(2003)]{tk_ladalada} Lada, C.~J., \& Lada, E.~A.\
2003, \araa, 41, 57

\bibitem[L{\'e}pine \& Bongiorno(2007)]{tk_lepine} L{\'e}pine, S., \&
Bongiorno, B.\ 2007, \aj, 133, 889

\bibitem[Makarov et al.(2008)]{tk_makarov} Makarov, V.~V.,
Zacharias, N., \& Hennessy, G.~S.\ 2008, \apj, 687, 566

\bibitem[Plummer(1911)]{tk_plummer} Plummer, H.~C.\ 1911, \mnras, 
71, 460 

\bibitem[Portegies Zwart et al.(2001)]{tk_starlab} Portegies 
Zwart, S.~F., McMillan, S.~L.~W., Hut, P., 
\& Makino, J.\ 2001, \mnras, 321, 199 

\bibitem[Poveda et al.(2007)]{tk_poveda2007} Poveda, A., Allen, C., \&
Hern{\'a}ndez-Alc{\'a}ntara, A.\ 2007, IAU Symposium, 240, 417

\bibitem[Quinn \& Smith(2009)]{tk_quinn} Quinn, D.~P., \& Smith,
M.~C.\ 2009, arXiv:0908.3640

\bibitem[Wasserman \& Weinberg(1991)]{tk_wasserman} Wasserman, I., \&
Weinberg, M.~D.\ 1991, \apj, 382, 149


\end{thebibliography}
\end{document}